\documentclass[journal,letterpaper]{IEEEtran}
\usepackage{amsmath,amssymb,amsthm}
\usepackage{microtype}
\usepackage{enumitem}

\usepackage{mathtools}

\usepackage{nth}

\usepackage{booktabs}
\usepackage{graphicx,xcolor}
\usepackage{hyperref}

\newtheorem{lemma}{Lemma}

\theoremstyle{definition}
\newtheorem{definition}{Definition}

\usepackage[utf8]{inputenc}    
\usepackage[T1]{fontenc}       

\theoremstyle{remark}
\newtheorem{remark}{Remark}

\newcommand{\bR}{\ensuremath\mathbb{R}}
\newcommand{\bx}{\ensuremath\boldsymbol x}
\newcommand{\bu}{\ensuremath\boldsymbol u}
\newcommand{\by}{\ensuremath\boldsymbol y}
\newcommand{\cX}{\ensuremath\mathcal{X}}
\newcommand{\cU}{\ensuremath\mathcal{U}}
\newcommand{\cY}{\ensuremath\mathcal{Y}}
\newcommand{\cL}{\ensuremath\mathcal{L}}

\long\def\Note#1{\bgroup\color{red}#1\egroup}

\makeatletter
\providecommand*{\diff}%
	{\@ifnextchar^{\DIfF}{\DIfF^{}}}
\def\DIfF^#1{%
	\mathop{\mathrm{\mathstrut d}}%
		\nolimits^{#1}\gobblespace}
\def\gobblespace{%
		\futurelet\diffarg\opspace}
\def\opspace{%
	\let\DiffSpace\!%
	\ifx\diffarg(%
		\let\DiffSpace\relax
	\else
		\ifx\diffarg[%
			\let\DiffSpace\relax
		\else
			\ifx\diffarg\{%
				\let\DiffSpace\relax
		\fi\fi\fi\DiffSpace}
\makeatother

\usepackage{tikz}
\usetikzlibrary{shapes,arrows}
\tikzstyle{block} = [draw, fill=blue!20, rectangle, 
    minimum height=2.5em, minimum width=5em]
\tikzstyle{sum} = [draw, fill=blue!20, circle, node distance=.8cm]
\tikzstyle{input} = [coordinate]
\tikzstyle{output} = [coordinate]
\tikzstyle{pinstyle} = [pin edge={to-,thin,black}]
\usetikzlibrary{positioning}
\usetikzlibrary{patterns}
\usetikzlibrary{calc}
\usetikzlibrary{shapes.misc}
\usetikzlibrary{backgrounds}

\usepackage{pgfplots}
\pgfplotsset{compat=newest}
\pgfplotsset{plot coordinates/math parser=false}

\usepackage[style=ieee,backend=biber,doi=false,isbn=false]{biblatex}
\usepackage{balance}\balance
\usepackage{algorithm2e}

\title{A Data-driven Adaptive Controller Reconfiguration for Fault Mitigation: A Passivity Approach}
\author{Hasan Zakeri\thanks{\quad}%
\thanks{Department of Electrical Engineering, University of Notre Dame, USA. \url{hzakeri@nd.edu}, \url{antsaklis.1@nd.edu}} and Panos J. Antsaklis
\thanks{The partial support of ARO under Grant No. ARL~W911NF-17-1-0072  is gratefully acknowledged.}\thanks{\quad }}

\begin{document}

\maketitle
\begin{abstract}
This paper presents a new data-driven fault identification and controller reconfiguration algorithm. The presented algorithm relies only on the system's input and output data, and it does not require a detailed system description. The proposed algorithm detects changes in the input-output behavior of the system, whether due to faults or malicious attacks and then reacts by reconfiguring the existing controller. This method does not identify the internal structure of the system nor the extent and nature of the attack; hence it can quickly react to faults and attacks. The proposed method can be readily applied to various applications without significant modifications or tuning, as demonstrated by the examples in the paper.
\end{abstract}

\section{Introduction}
Faults in a densely interconnected and tightly coupled network consisting of computation, communication and control components can lead to cascading effects disrupting the operation of the network. Modern dynamical systems should have the ability to restructure and adjust the control loop in the event of a fault. The nature and extent of faults in most cases are not predictable in advance, and considering all possibilities case by case in the design is not realistic. Designing a controller that is inherently fault-tolerant can also compromise the nominal performance. This study presents a data-driven method to detect a fault in the system through input and output data, without relying on a full system description, and then mitigating the effects of the fault through an adaptive controller reconfiguration. 

Several methods exist to improve the response of a system to a fault. \emph{Fault-tolerant control} deals with systems subjects to a fault (as opposed to classical control which only consider systems during normal operations). These methods are divided into ``passive'' approaches (not to be confused with passivity based approaches) and ``active'' approaches. In the first category, the controller can handle faults without any changes (this includes robust control approaches). The type and the extent of the faults they can tolerate is limited, and they are designed to address different fault cases and nominal operation, resulting in suboptimal overall performance. A formal definition and examples are reported in~\autocite{blanke_what_2000}. The active approach, on the other hand, refers to designing strategies where a ``re-design'' or ``reconfiguration'' happens in response to the occurrence of the fault. 
These approaches either explicitly isolate and identify the fault, or 
the controller changes in response to the fault, but without explicitly identifying the fault~\autocite{ioannou_robust_2012}. See also~\autocite{jiang_active_2003} for an example. See~\autocite{steffen_control_2005} for a review of different fault identification techniques. 

Reconfigurable control, which refers to re-design or re-adjustment of the control algorithm to ensure safe operation of the dynamical system with some performance guarantees in the event of a fault~\autocite{lunze_reconfigurable_2008}, has been studied via several approaches. \emph{Fault hiding control} places a reconfiguration block between the nominal controller and the faulty plant such that the system appears to be in a no-fault condition to the controller~\autocite{blanke_diagnosis_2016,rotondo_virtual_2014,blanke_diagnosis_2016}. Model matching is a methodology where a control system is designed to make the output of the plant to follow the output of the desired behavioral model~\autocite{gao_reconfigurable_1992}. The pseudo-inverse method (PIM), which is a widely used design approach for reconfigurable control systems, is a special case of classical linear model-following control~\autocite{gao_stability_1991}. We also refer readers to~\autocite{noura_fault-tolerant_2009,yen_online_2003,richter_reconfigurable_2011,staroswiecki_progressive_2006} and the references therein for other reconfiguration approaches. Recently, model predictive control reconfiguration and predictive based fault detection and reconfiguration have been studied as well~\autocite{esfahani_distributed_2016,zhou_reconfigurable_2016,burns_reconfigurable_2018,ferranti_fault-tolerant_2018}.

In this paper, we present a new fault identification method based on input and output data from the system. This method does not rely on a full detailed description of the system's behavior; therefore it can be readily applied to different applications. Since this method does not try to identify a detailed model for the system, it has the advantage of quick detection of the fault. Based on this method, we also present an adaptive controller reconfiguration to mitigate the effect of the fault. The controller reconfiguration method does not try to design a new controller from scratch, but it relies on the existing controller in the system (relying on the existing controller is an essential criterion in many industrial applications) and mitigates the fault by interfacing the controller with a real matrix which is determined online. In other words, an interface is wrapped around the current controller to mitigate the effects of the fault. In this paper, we focus on describing this novel approach in detail, explaining how it can be applied in practice.

After the preliminaries in \autoref{sec:pre}, we present the fault identification and mitigation method is \autoref{sec:main}. Even though the identification and control reconfiguration are presented as one algorithm, one can employ the identification process separately as well. \hyperref[sec:example]{Section~\ref*{sec:example}} presents examples of using this algorithm with simulation results. Conclusions and further directions are presented in \autoref{sec:conc}.

\section{Preliminaries}\label{sec:pre}
Here we introduce the passivity indices of a system, and then we cover stability results and a passivation method based on passivity indices.
Consider a  continuous-time dynamical system \(\mathbf H:\bu\to\by\), where \(\bu\in\cU\subseteq\bR^m\) denotes the input and \(\by\in\cY\subseteq\bR^p\) denotes the corresponding output. There exists a real-valued function \(w(\bu(t),\by(t))\) (often written as \(w(t)\) when clear from content) associated with \(\mathbf H\), such that for all input and output pairs of \(\bu(t)\) and \(\by(t)\) of the system,
\begin{equation}
	\int\limits_{t_0}^{t_1}\vert w(t)\vert\diff t<\infty.
\end{equation}
This function is called  \emph{supply rate function.} 
for every \(t_0\) and \(t_1\). Now consider a continuous-time system described by
\begin{equation}\label{eq:system}
	\begin{aligned}
		\dot \bx&=f(\bx,\bu)\\
		\by&=h(\bx,\bu),
	\end{aligned}
\end{equation}
where \(f(\cdot,\cdot)\) and \(h(\cdot,\cdot)\) are Lipschitz mappings of proper dimensions, and assume the origin is an equilibrium point of the system; i.e., \(f(0,0)=0\) and \(h(0,0)=0.\) 

\begin{definition}\label{def:dissipativity}
The system described by~\eqref{eq:system} is called \emph{dissipative with respect to supply rate function \(w(\bu(t),\by(t)),\)} if there exists a nonnegative real-valued scalar function \(V(\bx),\) called the \emph{storage function,} such that \(V(0)=0\) and for all \(\bx_0\in\cX,\) all \(t_1\geq t_0,\) and all \(\bu\in\bR^m\) 
\begin{equation}\label{eq:dissipativity}
	V(\bx(t_1))-V(\bx(t_0))\leq\int\limits_{t_0}^{t_1}w(\bu(t),\by(t))\diff t.
\end{equation}
where \(\bx(t_0)=x_0\) and \(\bx(t_1)\) is the state at \(t_1\) resulting from initial condition \(x_0\) and input function \(u(\cdot).\)
Inequality~\eqref{eq:dissipativity} is called \emph{dissipation inequality} and expresses the fact that the energy ``stored'' in the system at any time \(t\) is not more than the initially stored energy plus the total energy supplied to the system by its input during this time. 
\end{definition}
\begin{definition}\label{def:passivity}
The system~\eqref{eq:system} is called \emph{passive,} if it is dissipative with respect to the supply rate function \(w(\bu,\by)=\bu^\intercal\by.\) If \(V(\bx)\) is differentiable, then this is equivalent to 
\begin{equation}\label{eq:dissipativitydiff}
	\dot V(\bx)\triangleq\frac{\partial V}{\partial\bx}\cdot f(\bx,\bu)\leq \bu^\intercal\by.
\end{equation}
\end{definition}
Passivity indices are introduced as measures of passivity and they extend passivity based tools to non-passive systems as well.
\begin{definition}[Input Feed-forward Passivity Index]\label{def:IFP_index}
The system~\eqref{eq:system} is called \emph{Input Feed-forward Passive (IFP)} if it is dissipative with respect to supply rate function \(w(\bu,\by)=\bu^\intercal\by-\nu\bu^\intercal\bu\) for some \(\nu\in\bR,\) denoted as IFP(\(\nu\)). Input feed-forward passivity index for system~\eqref{eq:system} is the largest \(\nu\) for which the system is IFP. This is equivalent to the following dissipativity inequality
\begin{equation}
    \dot V(x)\leq u^\intercal y-\nu u^\intercal u
\end{equation}
holding for the largest \(\nu.\)
\end{definition}
IFP index is equivalent to the largest gain that can be put in a negative feed-forward interconnection with the system such that the overall system is passive.
\begin{definition}[Output Feedback Passivity]\label{def:OFP_index}
The system~\eqref{eq:system} is called \emph{Output Feedback Passive (OFP)} if it is dissipative with respect to supply rate function \(w(\bu,\by)=\bu^\intercal\by-\rho\by^\intercal\by\) for some \(\rho\in\bR,\) denoted as OFP(\(\rho\)). Output feedback passivity index for system~\eqref{eq:system} is the largest \(\rho\) for which the system is OFP.  This is equivalent to the following dissipativity inequality
\begin{equation}
    \dot V(x)\leq u^\intercal y-\rho y^\intercal y
\end{equation}
holding for the largest \(\rho.\)

\end{definition}
OFP index is the largest gain that can be placed in positive feedback with a system such that the interconnected system is passive. If either one of the indices for a system is positive, we say that the system has an ``excess of passivity,'' and similarly, if either one is negative, we say the system has a ``shortage of passivity.''

When applying the two indices simultaneously, a system is said to have IFP(\(\epsilon\)) and OFP(\(\delta\)), or IF-OFP(\(\epsilon,\delta\)), based on the following dissipation inequality:
\begin{equation}\label{eq:both_indices}
    \int\limits_0^T[(1+\epsilon\delta)\bu^\intercal\by-\delta\by^\intercal\by-\epsilon\bu^\intercal\bu]\diff t\geq V(\bx(T))-V(\bx(0)).
\end{equation}
When \(\epsilon=0\) and \(\delta=0\) the passivity index condition reduces to the definition of passivity~\autocite{process.passivity}.
\begin{remark}
When the dynamical system is given as an input-output map (with no internal description or initial conditions), it is common to assume the storage function is equivalent to zero; i.e., if the system is defined as \(G:u\mapsto y,\) then passivity is equivalent to \(u^\intercal Gu\geq0\) or \(u^\intercal y\geq0\) for all \(u\in\cU.\)
\end{remark}
Passivity indices under operationa limitations for nonlinear systems as well as approximate methods to find them are presented in~\autocite{zakeri_passivity_2019}. Local passivity indices for nonlinear systems and sum of squares methods to find the local indices are also presented in~\autocite{zakeri.acc2016}.
\subsection{Stability Properties}
\begin{lemma}[\autocite{Khalil}]\label{thm:indices}
Consider the feedback interconnection of \autoref{fig:feedback:general} and suppose each feedback component satisfies the inequality
\begin{equation}
    \dot V_i\leq\bu_i^\intercal\by_i-\nu_i\bu_i^\intercal\bu_i-\rho_i\by_i^\intercal\by_i,
\end{equation}
for some storage function \(V_i(\bx_i)\) where \(i=1,2.\) Then, the closed-loop map from \(\begin{bmatrix}r_1\\r_2\end{bmatrix}\) to \(\begin{bmatrix}\by_1\\\by_2\end{bmatrix}\) is finite gain \(\cL_2\) stable if \(\nu_1+\rho_2>0\) and \(\nu_2+\rho_1>0.\)
\end{lemma}
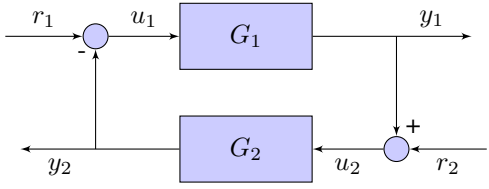
\begin{figure}
    \centering
    \begin{tikzpicture}[auto, node distance=.8cm,>=latex']
        \node [input, name=input] {};
        \node [sum, right of=input, node distance = 1.2cm] (sum) {};
        \node [block, right of=sum, 
                node distance=2cm] (system) {$G_1$};
        \node [output, right of=system, node distance=3cm] (output) {};
        \node [block, below of=system, node distance=1.5cm, name=system2] (measurements) {$G_2$};
        \node [sum, right of = measurements, node distance=2cm] (sum2) {};
        \node [input, right of = sum2, node distance = 1.2cm] (input1){};
        \node [output, left of = measurements, node distance = 3cm] (output2) {};
        
        \draw [draw,->] (input) -- node {$r_1$} (sum);
        \draw [->] (sum) -- node {$u_1$} (system);
        \draw [->] (system) -- node [name=y, pos=0.75] {$y_1$}(output);
        \draw [->] (output -| sum2) -- node[pos=0.9] {+} (sum2);
        \draw [->] (input1) -- node {$r_2$} (sum2);
        \draw [->] (sum2) -- node {$u_2$} (measurements);
        \draw [->] (measurements) -- node[pos=0.75] {$y_2$} (output2);
        \draw [->] (output2 -| sum) -- node[pos=0.95]{-} (sum);
    \end{tikzpicture}
    \caption{The general feedback interconnection of two dynamical systems \(G_1\) and \(G_2\)}
    \label{fig:feedback:general}
\end{figure}
The above theorem assumes the validity of the feedback interconnection, i.e., both systems are square with the same number of inputs and outputs. For a more general interconnection, see e.g.~\autocite{moylan_stability_1978,vidyasagar_nonlinear_2002}.

A relaxed version of the above theorem, stated for linear systems, is given as follows.
\begin{lemma}[\autocite{mccourt_dissipativity_2013,mccourt_passivity_2009}]\label{thm:stability:indices:1}
Consider the feedback interconnection of two linear systems where the indices exist for both systems. Assume that one of the two systems lacks either OFP or IFP on some time interval. This interconnection is Lyapunov stable if the following inequalities hold,
\begin{align}
    \rho_1+\nu_2&\geq0\\
    \rho_2+\nu_1&\geq0.
\end{align}
\end{lemma}
For more information on passivity indices and their applications in Cyber-physical system design, see~\autocite{survey,agarwal_passivity_2018}.
\begin{figure}[t]
    \centering
    \begin{tikzpicture}[auto, node distance=.8cm,>=latex']
        \node [input, name=input] {};
        \node [sum, right of=input, node distance = 1.2cm] (sum) {};
        \node [block, right of=sum, 
                node distance=2cm] (system) {$G$};
        \node [output, right of=system, node distance=3cm] (output) {};
        \node [block, below of=system, node distance=1.5cm, name=system2] (measurements) {$C$};
        \node [output, left of = measurements, node distance = 3cm] (output2) {};
        \draw [draw,->] (input) -- node {$r$} (sum);
        \draw [->] (sum) -- node {$e$} (system);
        \draw [->] (system) -- node [name=y, pos=0.75] {$y$}(output);
        \draw [->] ($(output)!0.4!(system)$) |- (measurements) ;
        \draw [->] (measurements) -|  node[pos=0.95]{-} (sum);
    \end{tikzpicture}
    \caption{The feedback interconnection of dynamical system \(G\) and controller \(C\) in nominal condition}
    \label{fig:feedback}
\end{figure}
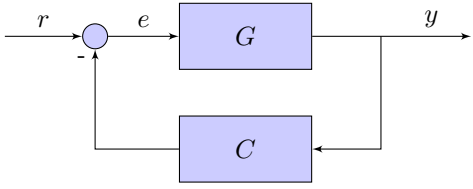

\subsection{Passivation and Design based on the M-matrix Method}
Passivity indices can be adjusted by series, feedback, or parallel interconnections. A generalization of these methods is given in~\autocite{xia_passivity_2014} by using an input-output transformation matrix. Appropriate design of this matrix, called \emph{the M-matrix,} guarantees positive passivity levels for the system. Consider the system \(G\) and a general input-output transformation matrix \(M.\)
 The matrix \(M\) is considered to be invertible and defined as 
\begin{equation}
    M\triangleq\begin{bmatrix}
        m_{11}I&m_{12}I\\m_{21}I&m_{22}I.
    \end{bmatrix}
\end{equation}
It is shown in~\cite{xia_passivity_2014} that the passivity indices of the system \(\Sigma_0:\bu_0\to\by_0\) defined as 
\begin{equation}
    \begin{bmatrix}u_0\\y_0\end{bmatrix}=M\begin{bmatrix}u\\y\end{bmatrix}
\end{equation}
depend on the gain \(\gamma\) of system \(G\) and the elements of \(M,\) as stated in \autoref{thm:Mmatrix}.
\begin{lemma}\label{thm:Mmatrix}
    Consider a finite gain stable system \(C\) with gain \(\gamma\) and a passivation matrix \(M\) as shown in \autoref{fig:M}. The system \(\Sigma_0:y\to u\) is 
    \begin{enumerate}
        \item passive, if \(M\) is chosen such that
        \begin{equation}\label{eq:M:passive}
            m_{11}=m_{21},\ m_{22}=-m_{21},\ m_{11}\geq m_{22}\gamma>0.
        \end{equation}
        \item OFP with OFP level \(\rho_0=\frac12\left(\frac{m_{11}}{m_{21}}+\frac{m_{12}}{m_{22}}\right)>0,\) if 
        \begin{equation}\label{eq:M:OFP}
            m_{21}\geq m_{22}\gamma>0,\quad m_{11}m_{22}>m_{12}m_{21}>0.
        \end{equation}
        \item IFP with IFP level \(\nu_0=\frac12\left(\frac{m_{21}}{m_{11}}+\frac{m_{22}}{m_{12}}\right)>0,\) if
        \begin{equation}\label{eq:M:IFP}
            m_{11}\geq m_{12}\gamma>0,\quad m_{12}m_{21}>m_{11}m_{22}>0.
        \end{equation}
        \item IF-OFP with passivity indices \(\delta_0=\frac12\frac{m_{11}}{m_{21}}>0\) and \(\epsilon_0=\frac a2\frac{m_{21}}{m_{11}}>0,\) if
        \begin{equation}\label{eq:M:IFOFP}
            m_{11}>0,\quad m_{12}=0,\quad m_{21}\geq\frac{m_{22}\gamma}{\sqrt{1-a}}>0,
        \end{equation}
        where \(0<a<1\) is an arbitrary real number.
    \end{enumerate}
\end{lemma}
For proof of \autoref{thm:Mmatrix}, see \autocite{xia_passivity_2014-1}. This passivation method is applied in~\autocite{xia_guaranteeing_2015} to a human controller. 

\section{Data-driven Algorithm}\label{sec:main}
Consider the usual control loop in \autoref{fig:feedback} as a special case of \autoref{fig:feedback:general}. Assume, under nominal operation, this loop satisfies the condition in \autoref{thm:indices}; i.e., if the system \(G\) has passivity indices \((\rho,\nu)\) and the controller has \((\rho_c,\nu_c),\) then \(\rho+\nu_c>0\) and \(\nu+\rho_c>0.\) 

The algorithm looks at the input and output data from the system, namely \(u(t)\) and \(y(t),\) then derives estimates \(\tilde\rho\) and \(\tilde\nu\) of the passivity indices. These two estimates are then compared to set thresholds to detect whether there have been any significant changes to the system or not. The thresholds should be chosen close to the real indices of the system. If at least one of the estimated indices is lower than the threshold, and it has not been lower before, then the controller reconfiguration procedure will be initiated. The reconfiguration procedure involves the \(M\)-matrix discussed earlier. The reconfiguration will vary based on how both indices compare to the thresholds. If they are both lower, then the design objective would be to compensate for both of them; otherwise the reconfiguration will compensate for only one of them. It is important to note that we do not try to change the passivity indices of the plant but to make sure that the loop satisfies \autoref{thm:indices}.
\begin{figure}[b]
    \centering
    \begin{tikzpicture}[auto, node distance=1cm,>=latex']
        \node [input, name=input] {};
        \node [sum, right of=input, node distance = 1.5cm] (sum) {};
        \node [block, right of=sum, minimum width=6em, minimum height = 3em,
                node distance=2.5cm] (system) {$G$};
        \node [output, right of=system, node distance=4cm] (output) {};
        \draw [draw,->] (input) -- node {$r$} (sum);
        \draw [->] (sum) -- node {$e$} (system);
        \draw [->] (system) -- node [name=y, pos=0.75] {$y$}(output);
        \node [input, name=dist, above = 1 cm of system] {};
        \draw [->] (dist) -- node {$d$} (system);
        
        \node [block, minimum width = 6cm, below = 1.5cm of system] (M) {M};
        \node [block, below = .75cm of M,
                 minimum width=6em, minimum height = 3em] (G) {\(C\)};
        \draw [->] (G.west) -| node {$u_0$}  (M.south west -| sum);
        \draw [->] ($(M.south west)!0.9!(M.south east)$) |- node {$y_0$} (G.east);
        \coordinate (F) at ($(M.north west)!0.9!(M.north east)$);
        \draw [->] (F |- system.east) -- (F);
        \draw [->] (M.north west -| sum) -- node[pos=0.9] {$-$} node {$u$}  (sum);
        \coordinate (A) at ($(M.south east)+(0.5,0)$);
        \coordinate (B) at ($(G.south east)+(0,-0.5)$);
        \begin{scope}[on background layer]
            \draw[rounded corners=3mm, dashed, fill=orange!30!white, label=Rec] ($(M.north west)+(-0.5,0.5)$) rectangle (B -| A);
            \node [above  = 0.5cm of M] {$\Sigma_0$};
        \end{scope}
    \end{tikzpicture}
    \caption{The  feedback interconnection of a dynamical system \(G\) and controller \(C\) with the reconfiguration interface}
    \label{fig:feedback2}
\end{figure}
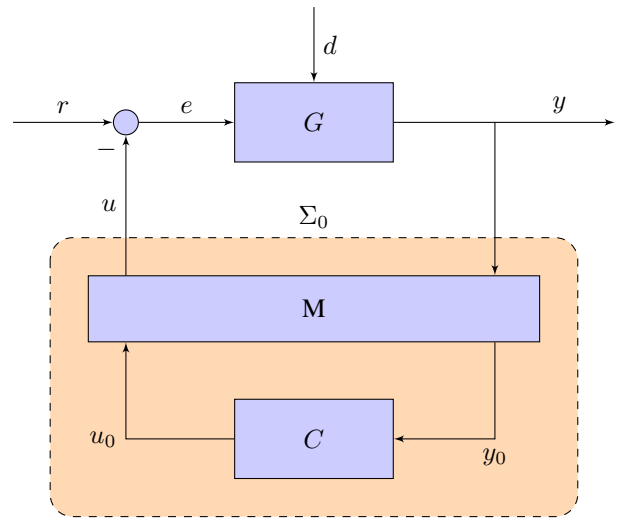
\begin{algorithm}
\caption{The identification and reconfiguration algorithm}\label{alg}
    \SetAlgoLined
    \KwData{Input and output to the system \(e\) and \(y\)}
    \KwResult{The reconfiguration matrix \(M\) }
    \(\rho_{\min}\gets\infty\)\;
    \(\nu_{\min}\gets\infty\)\;
    \(M\gets I_2\)\;
    \While{there is no manual override}{
        $t\gets$ time\;
        Compute \(\rho_0\) and \(\nu_0:\)
        \[\bar\rho=\frac{\int_0^te^\intercal y\diff\tau}{\int_0^ty^\intercal y\diff\tau},
        \quad
        \bar\nu_0=\frac{\int_0^te^\intercal y\diff\tau}{\int_0^te^\intercal e\diff\tau}
        \]
        \If{\(\bar\rho<\rho_0\) and \(\bar\nu\geq\nu_0\)}{
            Indicate a fault\;
            \eIf{\(\bar\rho<\rho_{\min}\)}{
                \(\rho_{\min}\gets\bar\rho\)\;
                Compensate for IFP\\
                \qquad using \autoref{thm:Mmatrix} and Eq.~\eqref{eq:M:IFP}\;
                \(\nu_c(new)+\bar\rho>\varepsilon\)\;
            }{
                Already compensated, move on\;
            }
        }
        \If{\(\bar\rho\geq\rho_0\) and \(\bar\nu<\nu_0\)}{
            Indicate a fault\;
            \eIf{\(\bar\nu<\nu_{\min}\)}{
                \(\nu_{\min}\gets\bar\nu\)\;
                Compensate for OFP\\
                \qquad using \autoref{thm:Mmatrix} and Eq.~\eqref{eq:M:OFP}\;
                \;
                \(\rho_c(new)+\bar\nu>\varepsilon\)\;
            }{
                Already compensated, move on\;
            }
        }
        \If{\(\bar\rho<\rho_0\) and \(\bar\nu<\nu_0\)}{
            Indicate a fault\;
            \eIf{\(\bar\nu<\nu_{\min}\) or \(\bar\rho<\rho_{\min}\)}{
                \(\nu_{\min}\gets\bar\nu\)\;
                \(\rho_{\min}\gets\bar\rho\)\;
                Compensate for IF-OFP\\
                \qquad using \autoref{thm:Mmatrix} and Eq.~\eqref{eq:M:IFOFP}\;
                \(\rho_c(new)+\bar\nu>\varepsilon\)\;
                \(\nu_c(new)+\bar\rho>\varepsilon\)\;
            }{
                Already compensated, move on\;
            }
        }
    }
\end{algorithm}

The algorithm is presented in \autoref{alg}. In \autoref{alg}, \(\bar\rho\) and \(\bar\nu\) are estimates of the indices, and the variables \(\rho_{\min}\) and \(\nu_{\min}\) keep track of changes in the estimated indices. If any of the indices go below these thresholds, the reconfiguration matrix needs to be recomputed. Otherwise, the index has had a lower value, and the corresponding reconfiguration is still valid. Parameters \(\rho_0\) and \(\nu_0\) are desired or safe values for the indices, and as long as the indices are not lower than these thresholds, the system is performing properly. Once \(\bar\rho\) and \(\bar\nu\) go below thresholds \(\rho_0\) and \(\nu_0,\) this indicates a malfunction in the system (more on this in the next section). 

\subsection{A Closer Look at the Estimation of the Passivity Indices}
It is important to note that our goal here is not to derive precise estimates of the passivity indices. To do so, one requires to either excite the system with various inputs to achieve an upper bound for the indices  (see~\autocite{zakeri_passivity_2018,wu_experimentally_2013} for a detailed discussion), or find the right kind of input that will lead to the actual indices or close to the actual indices (see~\autocite{romer_ideas_2018,romer_sampling_2017,tanemura_efficient_2018} for example). However, both of these techniques are more suitable as offline methods. In the real-time estimation of the indices, one usually does not have the freedom to modify the input applied to the system. 

The estimations can be represented as functions of time
\begin{equation}\label{eq:integration}
    \bar\rho(t)=\frac{\int_0^te^\intercal y\diff\tau}{\int_0^ty^\intercal y\diff\tau},
    \qquad
     \bar\nu_0(t)=\frac{\int_0^te^\intercal y\diff\tau}{\int_0^te^\intercal e\diff\tau}.
\end{equation}
If \(\rho\) and \(\nu\) represent the actual indices of the system, based on the definition, the dissipation inequality should hold for all possible inputs. However, \(\bar\rho\) and \(\bar\nu\) correspond to a limited set of inputs to the system; therefore, we can bound them as
\begin{equation}\label{eq:ineq}
    \bar\rho(t)\geq\rho,\ \bar\nu(t)\geq\nu,\quad\forall t,u(t).
\end{equation}
 This is mainly because the definition of the indices implies a min-max optimization as
\begin{equation}
    \rho=\min_{u}\max_e{\epsilon}
\end{equation}
such that \(u^\intercal y-\epsilon y^\intercal y\) holds (\(\nu\) would be analogous). Therefore, for a particular input, our estimate is going to be greater or equal to the actual index. Here, in the proposed algorithm, the objective is not to precisely measure the system's passivity indices; rather, we use the estimates \(\bar\rho\) and \(\bar\nu\) as indicators of the system's operation: if they do not meet certain criteria, then it is an indication of a malfunction in the system. 

\paragraph*{Integral Saturation and Overuse}
If the system is running for a long time, there is a possibility that the integrators in~\eqref{eq:integration} saturate overtime. Another possibility in longtime use is that \(\int_0^tu^\intercal y\diff\tau,\) \(\int_0^ty^\intercal y\diff\tau,\) and \(\int_0^tu^\intercal u\diff\tau\) might become very large, and a fault in the system would not be reflected by changes of a passivity index. One possible remedy is to either reset the integrators once in a while or integrate over a moving window; i.e., the lower limit of the integrals in~\eqref{eq:integration} would be \(t-t_0.\)
\begin{figure}[b]
    \centering
    \newlength{\figwidth}
    \setlength{\figwidth}{0.8\linewidth}
    \input{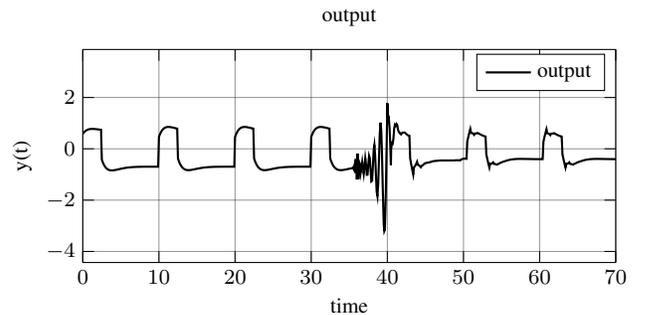}
    \caption{The output of the system with the data-driven algorithm in action when the fault is modeled as a time delay}
    \label{fig:io}
\end{figure}

\section{Examples}\label{sec:example}
In this section, we provide three different examples to demonstrate how this algorithm works. In each example, a nominal system is working in the loop with a controller. After sometime, a fault or attack happens in the system, which is modeled as a change in the dynamics of the system. The nominal dynamics, the nature of the change, and the new dynamics are unknown to the algorithm. In each example, the same algorithm is applied, with different parameters, to keep the closed-loop stable.\pagebreak[3] 

The first two examples are identical in the nominal operation and algorithm parameters. The same algorithm can handle different faults or attacks happening to the system. The first\pagebreak[3]  fault is modeled as a time-varying delay introduced to the system, and the second one is addition of nonlinear dynamics.\pagebreak[3]  It is worth noting that there are many methods to handle time-varying delays, and many methods to handle nonlinearities; however, they all need to be designed in advance for delay or nonlinearity, and they may compromise the design by being robust to the fault. In our presented method, there is no knowledge of what will happen in the system, and the alforithm will only reconfigure the controller if the nominal operation is compromised. The last example demonstrates a physical system with mass, damper, and spring, where the spring changes dynamics overtime.

\subsection{Time Delay due to Actuator Failure or DoS Attacks}
This numerical example demonstrate the effectiveness of the proposed method in identifying and then mitigating a malfunction where the fault is a time delay, which can model either an actuator fault or a Denial of Service (DoS) attack~\autocite{huang_understanding_2009}. The system is running in nominal operation with its own controller in the loop. A delay is introduced in the system, and is gradually increased. 

Consider the loop configuration of \autoref{fig:feedback}. The nominal system is given as 
\begin{equation}\label{eq:sys.ex1}
    G(s)=\frac{s^2+3s+2}{s^2+s+2}
\end{equation}
and is controlled by a lead compensator 
\begin{equation}\label{eq:c.ex1}
    C(s)=1.37\frac{s+0.91}{s+1.08}
\end{equation}
to satisfy some tracking objective. Under normal operation, the system is stable with adequate performance. 

A time-varying input delay \(\tau(t),\) modeling a DoS attack or actuator failure, is introduced to the system starting at time \(t=35.\) The delay gradually increases to reach the value of \(0.5s\) at \(t=40.\) With no further fault tolerant control in action, the loop will become unstable (the diverging output is not depicted here).

\begin{figure}[b]
    \includegraphics[width=.9\linewidth]{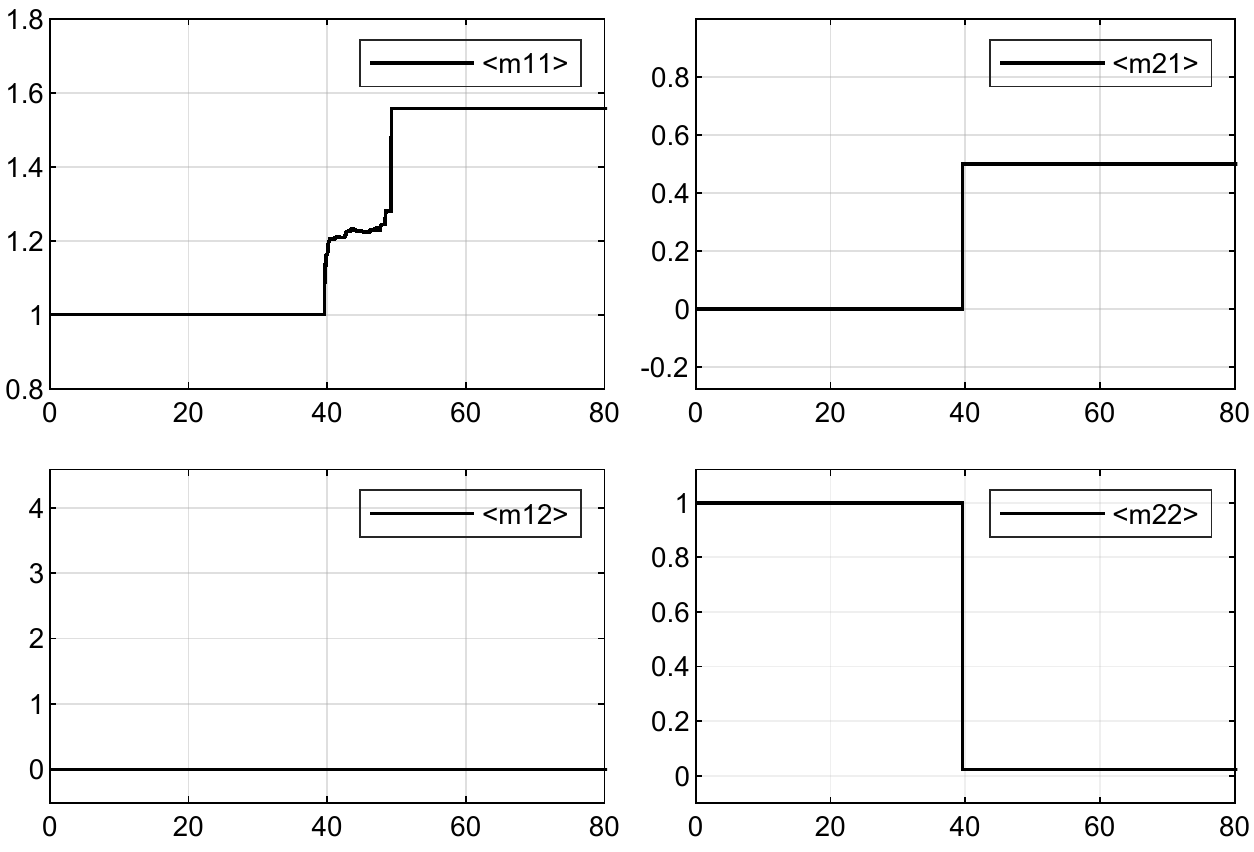}
    \caption{The elements of the reconfiguration matrix \(M\) to mitigate the effect of time delay}
    \label{fig:M}
\end{figure}

The proposed data-driven algorithm detects the fault in the system and reconfigure the controller in time by online design of the \(M\) matrix. At each time step, a new estimate is done, and if necessary, the reconfiguration matrix is recomputed. The algorithm keeps the loop stable, as shown by the output of the system in \autoref{fig:io}. The elements of the \(M\)-matrix are also depicted in \autoref{fig:M}. It can be seen that because of the freedom in designing \(M,\) some elements are chosen as constant and change based on the measurements. As can be seen in \autoref{fig:io}, the system output starts to diverge after \(t=35,\) however the algorithm takes over around \(t=40\) and makes the system stable. The performance may not be ideal, but the role of the algorithm is to maintain stability quickly and with minimum knowledge of the system.

To achieve a better performance, after the reconfiguration has happened, the elements of the \(M\)-matrix can be slightly ajusted by a separate algorithm to improve the performance. Similar approaches are presented in~\autocite{xia_performance_2016,rahnama_passivation_2016} and will not be covered here.

It is worth noting that the algorithm did not include the transfer function of either the system or the controller, nor is it based on the nature and specifics of the fault.

\subsection{Nonlinear Dynamics}
\begin{figure}[!t]
    \centering\small\vskip-5mm
    \input{ex2fig1.tikz}
    \caption{Input and output of the closed-loop system when the system becomes nonlinear at \(t=40s\) and the algorithm is in operation}
    \label{fig:ex2.inout}
    \vspace{5mm}
    \includegraphics[width=.85\linewidth]{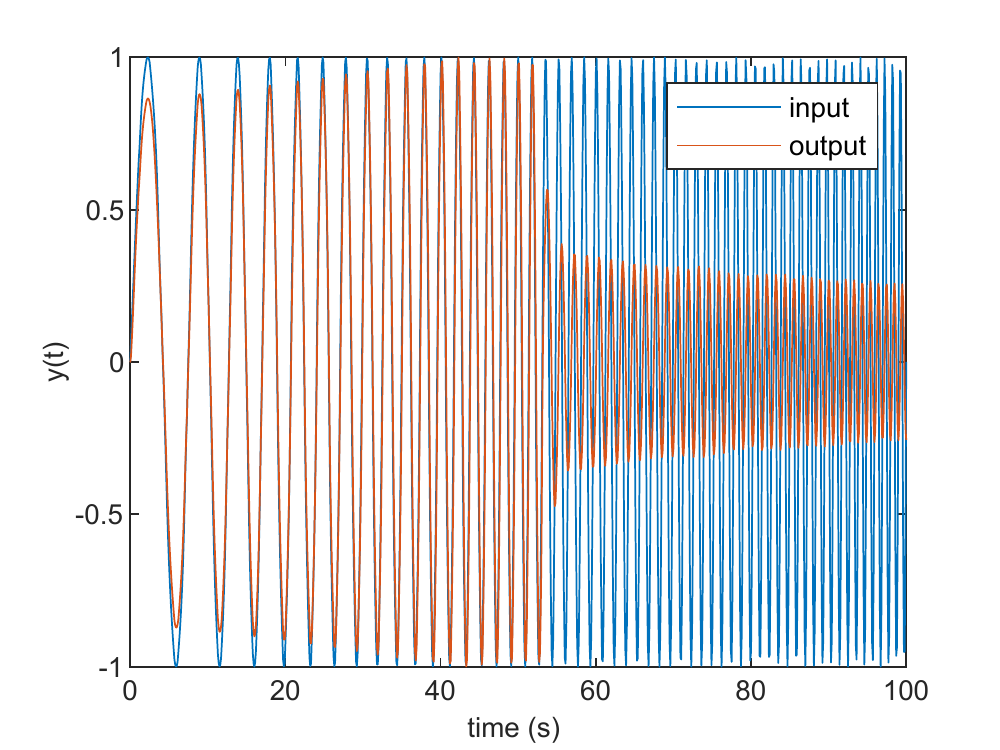}
    \caption{Input and output of the closed loop system involving a mass, damper, and spring system when the spring starts to soften at \(t=40s.\) The reconfiguration algorithm is in action and maintains stability of the loop.}
    \label{fig:ex3}
\end{figure}

Consider the system in~\eqref{eq:sys.ex1} where now the fault is a change in the dynamics; namely the linear dynamics will suddenly become nonlinear (next example will illustrate a physical example of such a change). Also consider the same compensator in the loop. The system~\eqref{eq:sys.ex1} can be written in state space form as 
\begin{equation}
    \begin{aligned}
        \dot x_1&=-x_1-2x_2+2u\\
        \dot x_2&=x_1\\
        y&=x_1+u.
    \end{aligned}
\end{equation}
At \(t=40s,\) a fault happens in the system and the new dynamics will be
\begin{equation}
    \begin{aligned}
        \dot x_1&=-x_1-2x_2+2u\\
        \dot x_2&=x_1-0.5x_2^2\\
        y&=x_1+u.
    \end{aligned}
\end{equation}
which has the same linearization. The compensator will fail to keep the system stable; however, the proposed algorithm, with the same exact parameters and thresholds as the last case, maintains the stability of the loop with suitable performace. \autoref{fig:ex2.inout} depicts the input and output of the closed-loop system when the algorithm is in operation. Without the fault mitigation algorithm, the system will be unstable. 

\subsection{Softening Spring}
For an example of a physical system, consider a base-excited mass damper spring system. If there is an ideal linear spring in the system, the equation of motion can be written as 
\begin{equation}
    m(\ddot y-\ddot u)+c(\dot y-\dot u)+k(y-u)=0
\end{equation}
where \(u(t)\) is the excitation displacement applied at the bottom and \(y(t)\) is the movement of the mass at the top (these equations have applications in active suspension systems). Overtime, the spring can gradually turn into a softening spring, modeled as 
\begin{equation}
    m(\ddot y-\ddot u)+c(\dot y-\dot u)+k(y-u)+\alpha(y-u)^3=0
\end{equation}
with \(\alpha<0.\)

In this example, the linear system (with \(\alpha=0\)) is working with a lag compensator in the loop given as 
\[
    C=4.8\frac{s+3.006}{s+2.485}
\]
with parameters \(m=2,c=3,\) and \(k=10.\) At \(t=40,\) the spring starts to soften, with \(\alpha\) gradually reaching \(-1\) at \(t=50.\) The controller is not able to keep the system stable.

The \autoref{alg} is applied to the loop with \(\rho_0=\nu_0=-0.15\) and detects the change and maintains the stability of the system. The parameters \(\nu_0\) and \(\rho_0\) are chosen as lower bounds for passivity indices of the nominal system.
The input and output of the closed-loop system is depicted in \autoref{fig:ex3}. Even though the system does not have perfect tracking after the fault, the system remains stable and operational. It is a trade-off between performance and maintaining safety and stability of the system under unknown malfunctions and faults. 
\begin{remark}
There is a degree of freedom in the elements of the \(M\)-matrix, and it can be exploited to improve performance. See for example~\autocite{rahnama_passivation_2016} for a bisimulation based performance optimization. The direct relation of elements of \(M\)-matrix and different performance criteria is still an open question.
\end{remark}

\section{Conclusion}\label{sec:conc}
In this paper, we presented some initial results of a new fault detection and adaptive controller reconfiguration algorithm. This method is based on the concept of passivity indices and relies only on the available data from input and output of the system. The presented method can detect changes in the system's behavior due to many factors including actuator faults and malicious attacks and can mitigate the fault or attacks by wrapping a reconfiguration matrix around the existing controller. The algorithm has shown significant potential as demonstrated by different examples.

\printbibliography

\end{document}